\documentclass[pra,showpacs,showkeys]{revtex4}
\usepackage{graphicx,amsmath,amsfonts,amssymb}
\usepackage{times}
\begin{document}
\title{Effect of weak measurement on entanglement distribution over noisy channels}
\author{Xin-Wen Wang,$^{1,2,}$\footnote{xwwang@mail.bnu.edu.cn} Sixia Yu,$^{2}$\footnote{cqtys@nus.edu.sg}
  Deng-Yu Zhang,$^{1}$ and C. H. Oh$^{2,}$\footnote{phyohch@nus.edu.sg}}
 \affiliation{$^1$College of Physics and Electronic Engineering, Hengyang Normal University, Hengyang 421002, China\\
  $^2$Center for Quantum Technologies, National University of Singapore, 2 Science Drive 3, Singapore 117542}

\begin{abstract}
Being able to implement effective entanglement distribution in noisy
environments is a key step towards practical quantum communication,
and long-term efforts have been made on the development of it.
Recently, it has been found that the null-result weak measurement
(NRWM) can be used to enhance probabilistically the entanglement of
a single copy of amplitude-damped entangled state. This paper
investigates remote distributions of bipartite and multipartite
entangled states in the amplitude-damping environment by combining
NRWMs and entanglement distillation protocols (EDPs). We show that
the NRWM has no positive effect on the distribution of bipartite
maximally entangled states and multipartite
Greenberger-Horne-Zeilinger states, although it is able to increase
the amount of entanglement of each source state (noisy entangled
state) of EDPs with a certain probability. However, we find that the
NRWM would contribute to remote distributions of multipartite W
states. We demonstrate that the NRWM can not only reduce the
fidelity thresholds for distillability of decohered W states, but
also raise the distillation efficiencies of W states. Our results
suggest a  new idea for quantifying the ability of a local filtering
operation in protecting entanglement from decoherence.
\end{abstract}



\maketitle

\section{Introduction}

It is well known that establishment of quantum entanglement among
distant parties is a prerequisite for many quantum information
protocols. Moreover, a necessary condition for perfectly
implementing these tasks is that the shared entangled states among
the users are maximally entangled pure states. In practice, however,
unavoidable interactions of the entangled systems with environments
during their distributions or storages would result in degradation
of the entanglement among the users. In other words, the
entanglement resources actually available are usually entangled
mixed states, which would decrease the fidelities and efficiencies
of quantum information processes.

To accomplish the aforementioned quantum information processing
tasks, the communicators need to transform the noisy entangled
states into maximally entangled pure states in advance. This raises
a problem which is also of theoretical interest: How can maximally
entangled pure states be extracted from shared entangled mixed
states by local operations? One solution, at least in principle, is
to use entanglement distillation protocols (EDPs) which function as
distilling a small number of entangled pure or nearly pure states
from a large number of entangled mixed states
\cite{76PRL722,77PRL2818,423N417,443N838,111PRL020502}. This means
perfect or nearly perfect entanglement-based quantum information
processing would be possible even in noisy environments by utilizing
the idea of entanglement purification.

However, the EDPs do not work for the inseparable states whose
fidelities or singlet fractions (which quantify how close the states
are to maximally entangled states \cite{76PRL722,54PRA3824}) are
less than some thresholds (e.g., 1/2 for two-qubit states
\cite{76PRL722,77PRL2818}), except that they have some special forms
or are hyperentangled
\cite{54PRA3824,66PRA022309,69PRA012308,45JPA125303,81PRA032307,5SR7815}.
Fortunately, Gisin \cite{210PLA151} discovered that the amount of
entanglement of an entangled mixed state could be raised
probabilistically by local filtering operations, which has been
proven in the experiment \cite{409N1014}. Moreover, local filtering
could be used to make trace-preserving local operations assisted by
classical communication so as to increase limitedly the fidelities
of some low-fidelity entangled mixed states with entanglement
unchanged
\cite{90PRL097901,65PRA022302,62PRA012311,90PRA052304,86PRA020304}.
These findings enable the entanglement of little-entangled particles
(even with fidelities less than the thresholds) to be distillable,
because they can be put through local filters, such that their
fidelities are over the related thresholds, prior to being subjected
to EDPs \cite{78PRL574}.

Recently, purification of a single-copy entangled mixed state by
local filtering operations has attracted considerable interest
\cite{82PRA052323,8NP117,86PRA052322,86PRA012325,86PRA032304,
350AP50,23CPB020304,89PRA022318,106EPL60003,89PRA062320,45JPA415303,321OC205,
64PRA010101,83PRL2656,77PRA012332,65PRA052318,96PRL220505,65PRA052319,82PRA022324,5SR8575},
due to the fact that it does not involve multiparticle collective
operations on multiple copies of source states and thus may reduce
the experimental difficulty, as well as can act as a complement to
entanglement distillation. The null-result weak measurement (NRWM, a
local filtering operation)\cite{89PRL127901} is widely used to
enhance the entanglement of various decohered states in
amplitude-damping (AD) or generalized AD environments
\cite{82PRA052323,8NP117,86PRA052322,86PRA012325,350AP50,23CPB020304,86PRA032304,89PRA022318,106EPL60003}.
The experimental viability of implementing a NRWM and its reversal
\cite{8NP117,97PRL166805,101PRL200401,19OE16309,111PRL180501,86PRA012333,110PRL070403,81PRA040103}
indeed makes it an elegant approach to protecting entanglement.
However, the filtering method cannot be applied for the direct
production of entangled pure states \cite{81PRL3279,81PRL2839}. To
obtain maximally entangled pure states for perfect remote quantum
information processing, EDPs are required. Then, a question arises,
namely, is the NRWM beneficial to entanglement distribution among
distant parties in terms of the efficiency of extracting maximally
entangled states, although it can improve with a certain probability
the entanglement of each source state (initial noisy entangled
state) of the EDP? This paper is addressing such an issue.

We consider entanglement distribution over AD channels. The aim of
the users is to share maximally entangled states. As mentioned
before, to achieve remote distribution of maximally entangled
states, we resort to the entanglement distillation. In previous
literatures \cite{82PRA052323,8NP117,86PRA052322,86PRA012325}, the
NRWM was introduced to raise the amount of entanglement of a
single-copy decohered state in AD environments. We here investigate
the impact of the NRWM on entanglement distribution efficiencies
(i.e., the efficiencies of distilling maximally entangled states) by
using it to enhance the entanglement of each decohered state before
starting the distillation procedures. We show that NRWMs would
decrease distillation efficiencies of bipartite maximally entangled
states and multipartite Greenberger-Horne-Zeilinger (GHZ) states
\cite{58AJP1131}. The efficiency (also known as yield in literature)
of an EDP is conventionally defined as the ratio of the number of
obtained maximally entangled states to that of source states
(inputs). Multipartite W-state \cite{62PRA062314} distribution,
however, exhibits different behaviors and features. That is to say,
the NRWM would contribute to increasing the efficiency of W-state
distribution with the existing EDP or its generalization and
reducing the fidelity threshold for distillability of the decohered
W state. Our results indicate that the NRWM is not necessarily
helpful to practical entanglement distributions, although it is able
to increase the amount of entanglement of a single-copy noisy
entangled state, and thus suggest a new approach to quantify the
ability of a local filtering operation in protecting entanglement
from decoherence.

The rest of this paper is organized as follows. We demonstrate the
uselessness of the NRWM to bipartite entanglement distribution in
Sec.~II, and discuss the effect of the NRWM on multipartite
entanglement distribution in Sec.~III. Finally, a summary is given
in Sec.~IV.

\section{Bipartite entanglement distribution}

The quantum channel considered in this paper is the AD channel. AD
decoherence is applicable to many practical qubit systems, including
vacuum-single-photon qubit with photon loss, photon-polarization
qubit traveling through a polarizing optical fiber or a set of glass
plates oriented at the Brewster angle, atomic qubit with spontaneous
decay, and superconducting qubit with zero-temperature energy
relaxation. The action of the AD channel on a qubit $l$ can be
described by two Krauss operators \cite{Nielsen}
\begin{eqnarray}
 && K_0^{(l)}=|0\rangle\langle 0|+\sqrt{\bar{d}_l}|1\rangle\langle1|,\nonumber\\
 && K_1^{(l)}=\sqrt{d_l}|0\rangle\langle 1|,
\end{eqnarray}
where $d_l$ stands for the damping rate satisfying $0\leqslant
d_l<1$ and $\bar{d}_l=1-d_l$. The AD channel is trace preserving,
that is, $\sum_{j=0,1}K_j^{(l)\dag}K_j^{(l)}=I$. Note that $d=0$
denotes the noise-free case, and it will not be considered in the
following context.

Assume the initial entangled state to be distributed to Alice and
Bob is a 2-qubit Bell state given by
\begin{equation}
\label{Bell}
  |\psi\rangle=\frac{1}{\sqrt{2}}(|01\rangle+|10\rangle).
\end{equation}
During the process of distributing or storing, the two qubits would
experience AD decoherence with decoherence strength $d_1$ and $d_2$,
respectively. The original entangled pure state then degrades into a
mixed state
\begin{eqnarray}
\label{rho}
 \rho_d&=&\sum\limits_{i,j=0,1}K_i^{(1)}\otimes K_j^{(2)}|\psi\rangle\langle\psi|
           K_i^{(1)\dag}\otimes K_j^{(2)\dag}\nonumber\\
&=&\frac{1}{2}\left(\sqrt{\bar{d}_2}|01\rangle+\sqrt{\bar{d}_1}|10\rangle\right)\left(\sqrt{\bar{d}_2}\langle01|
     +\sqrt{\bar{d}_1}\langle10|\right)+\frac{1}{2}(d_1+d_2)|00\rangle\langle00|,
\end{eqnarray}
where the superscripts of $K_{i,j}$ denote the qubit indices. The
concurrence (a universal entanglement measure for 2-qubit states
\cite{80PRL2245}) of $\rho_d$ is
\begin{equation}
 C(\rho_d)=\sqrt{\bar{d}_1\bar{d}_2}.
\end{equation}

As claimed and demonstrated in recently reports
\cite{82PRA052323,8NP117,86PRA052322,86PRA012325}, the concurrence
of the decohered state $\rho_d$ can be improved probabilistically by
performing locally each qubit a weak measurement, accompanied by a
bit flip operation before and after the weak measurement,
respectively. The weak measurement is a kind of measurement that
does not totally collapse the measured system. Practically, the weak
measurement on a qubit can be done by monitoring its environment
using a detector
\cite{8NP117,97PRL166805,101PRL200401,19OE16309,111PRL180501,86PRA012333,110PRL070403,81PRA040103}.
Whenever the detector registers an ``excitation'', one knows that
the qubit has totally collapsed into its ground state; if, however,
there is no ``excitation'' (null result), one knows that the qubit
state is just renormalized. Mathematically, such a measurement can
be described by two positive operators
\begin{eqnarray}
&& M_0=\sqrt{1-w}|1\rangle\langle1|+|0\rangle\langle0|,\nonumber\\
&& M_1=\sqrt{w}|1\rangle\langle1|.
\end{eqnarray}
If we discard the outcome of $M_1$, then $M_0$ denotes the NRWM
(null-result weak measurement) of strength $w$ $(0\leqslant w<1)$,
that partially collapses the system to the ground state. The NRWM in
fact uses post-selection to selectively map the state of a qubit. If
no outcome is discarded, the two operators $M_1$ and $M_0$ will
describe a noisy effect. Considering that a flip operation
$\sigma^x$ (conventional Pauli operator) is preformed on the system
before and after the NRWM $M_0$, respectively, the total process can
be described by the operator
\begin{eqnarray}
\label{M}
 M_w&=&\sigma^xM_0\sigma^x\nonumber\\
  &=&\sqrt{\bar{w}}|0\rangle\langle0|+|1\rangle\langle1|,
\end{eqnarray}
where $\bar{w}=1-w$. For convenience, $M_w$ will be directly
referred to as the NRWM operator. After Alice (holds the first
qubit) and Bob (holds the second qubit) performing NRWMs of strength
$w_1$ and $w_2$ on the entangled pairs, respectively, the state
$\rho_d$ becomes
\begin{eqnarray}
  \rho_w&=&\frac{1}{P_w}M_{w_1}\otimes M_{w_2}\rho_d M^{\dag}_{w_1}\otimes M^{\dag}_{w_2}\nonumber\\
   &=&\frac{\left(\sqrt{\bar{d}_2\bar{w}_1}|01\rangle+\sqrt{\bar{d}_1\bar{w}_2}|10\rangle\right)
        \left(\sqrt{\bar{d}_2\bar{w}_1}\langle01|+\sqrt{\bar{d}_1\bar{w}_2}\langle10|\right)
        +(d_1+d_2)\bar{w}_1\bar{w}_2|00\rangle\langle00|}
    {(d_1+d_2)\bar{w}_1\bar{w}_2+\bar{d}_2\bar{w}_1+\bar{d}_1\bar{w}_2},
\end{eqnarray}
where $P_w$ is the probability of getting the outcome of
$M_{w_1}\otimes M_{w_2}$, i.e. the probability of successful event,
given by
\begin{eqnarray}
\label{Pw}
 P_w&=&\mathrm{Tr}\left(M^{\dag}_{w_1}M_{w_1}\otimes M^{\dag}_{w_2} M_{w_2}\rho_d
 \right)\nonumber\\
 &=&\frac{1}{2}(d_1+d_2)\bar{w}_1\bar{w}_2+\frac{1}{2}\bar{d}_2\bar{w}_1+
         \frac{1}{2}\bar{d}_1\bar{w}_2.
\end{eqnarray}
Evidently, $\rho_w$ is equivalent to $\rho_d$ for $w_1=w_2=0$ that
means no weak measurement is made. Naturally, $P_w$ is then equal to
1. The concurrence of $\rho_w$ can be calculated as
\begin{equation}
 C(\rho_w)=\frac{\sqrt{\bar{d}_1\bar{d}_2\bar{w}_1\bar{w}_2}}{P_w}.
\end{equation}
$C(\rho_w)$ is larger than $C(\rho_d)$ provided that
$\sqrt{\bar{w}_1\bar{w}_2}>P_w$. Such a condition can be satisfied
for any $d_1$ and $d_2$ by choosing suitable $w_1$ and $w_2$. For
instance, the inequality always holds for $w_1=w_2$. It is easy to
see that when $C(\rho_w)\rightarrow 1$ (corresponding to
$w\rightarrow 1$), the success probability $P_w\rightarrow 0$.

Although the entanglement established between Alice and Bob was
improved by NRWMs, the shared entangled state is still not a
maximally entangled pure state that is a prerequisite for some
perfect quantum communications (e.g., teleportation). As mentioned
before, the filtering operations cannot be, even in principle,
applied for the direct production of maximally entangled states
\cite{81PRL3279,81PRL2839}. To obtain maximally entangled states,
Alice and Bob need further to utilize EDPs.

Next, we investigate whether the NRWM can help Alice and Bob to
raise the efficiency of getting maximally entangled states by
transforming the decohered state $\rho_d$ to $\rho_w$ using NRWMs
before starting the EDP. We will employ two EDPs, both of which
enable bipartite maximally entangled pure states to be extracted
from finite copies of $\rho_w$ or $\rho_d$ (corresponding to
$w_1=w_2=0$ in $\rho_w$). The first EDP will be called a two-copy
EDP, because each round of distillation only involves two copies of
input states \cite{54PRA3824}. The second EDP will be referred to as
a bisection EDP, because each round of distillation except the first
round divides the pairs of qubits into two blocks of equal length
\cite{80PRA014303}. The bisection EDP is up to now the most
efficient theoretical scheme for the amplitude-damped state $\rho_d$
or $\rho_w$ \cite{80PRA014303}, although it is much more difficult
than the two-copy EDP in the experiment.

\subsection{Two-copy EDP}

Suppose there is a collection of groups of source entangled pairs
$\rho_w$. Each group contains two pairs, one as the control pair and
the other as the target pair. Each party of Alice and Bob holds one
qubit of each pair. The EDP works as follows: (i) Alice and Bob
apply, respectively, a local controlled-not (CNOT) gate between the
two pairs of each group (i.e., the bilateral CNOT operation
\cite{54PRA3824}), where the control pair comprises the two control
qubits and the target one the two target qubits; (ii) they measure
locally the target pair in the computational basis
$\{|0\rangle,|1\rangle\}$; (iii) they keep the control pair if they
get the outcomes ``11'' (this means the success of extracting a
maximally entangled state) and ``00'' (in this case, the control
pair can be used for the second round of distillation), and discard
it otherwise.

It can be easily verified that if the outcome of this measurement on
a given target pair is ``11'', then the corresponding control pair
is left in the Bell state $|\psi\rangle$ which can be used for
faithful teleportation, etc. The probability of this event is
\begin{equation}
  P_{1}=\frac{2\bar{d}_1\bar{d}_2\bar{w}_1\bar{w}_2}{\left[(d_1+d_2)\bar{w}_1\bar{w}_2
    +\bar{d}_2\bar{w}_1+\bar{d}_1\bar{w}_2\right]^2}.
\end{equation}
Since each target pair has to be sacrificed for the measurement, the
yield from this procedure is $Y_{r_1}=P_{1}/2$. As for the
measurement outcome ``00'' of the target pair, the corresponding
control pair is left in the state
\begin{eqnarray}
 \rho_{r_1}=\frac{\left(\bar{d}_2\bar{w}_1|01\rangle+\bar{d}_1\bar{w}_2|10\rangle\right)
        \left(\bar{d}_2\bar{w}_1\langle01|+\bar{d}_1\bar{w}_2\langle10|\right)
        +(d_1+d_2)^2(\bar{w}_1\bar{w}_2)^2|00\rangle\langle00|}
    {(d_1+d_2)^2(\bar{w}_1\bar{w}_2)^2+\left(\bar{d}_2\bar{w}_1\right)^2
       +\left(\bar{d}_1\bar{w}_2\right)^2}.
\end{eqnarray}
The probability of this event is
\begin{equation}
 P_0=\frac{(d_1+d_2)^2(\bar{w}_1\bar{w}_2)^2+\left(\bar{d}_2\bar{w}_1\right)^2
       +\left(\bar{d}_1\bar{w}_2\right)^2}{\left[(d_1+d_2)\bar{w}_1\bar{w}_2+\bar{d}_2\bar{w}_1
       +\bar{d}_1\bar{w}_2\right]^2}.
\end{equation}
Evidently, two copies of $\rho_{r_1}$ can be used for the second
round of distillation following the procedure above. Then after $m$
rounds of distillation procedure, the efficiency (total yield) of
this EDP becomes
\begin{eqnarray}
 &&E'_f(\rho_w)=Y_{r_1}+Y_{r_2}+\cdots+Y_{r_m},\\
 &&Y_{r_1}=\frac{P_1}{2},\nonumber\\
 &&Y_{r_m}=\frac{Y_{r_{m-1}}\left(\bar{d}_1\bar{d}_2\bar{w}_1\bar{w}_2\right)^{2^{m-2}}}
    {2\left\{\left[(d_1+d_2)\bar{w}_1\bar{w}_2\right]^{2^{m-1}}
    +(\bar{d}_2\bar{w}_1)^{2^{m-1}}+(\bar{d}_1\bar{w}_2)^{2^{m-1}}\right\}},~~(m>1).\nonumber
\end{eqnarray}
Naturally, for a given EDP, the more entangled the source states
are, the higher efficiency would be obtained. As a consequence, the
value of $E'_f$ in a general case (i.e., $w_1$ and $w_2$ are not
simultaneously equal to zero) can always be larger than that of it
in the case $w_1=w_2=0$ for given $d_1$ and $d_2$, because the
source state $\rho_w$ can be more entangled than $\rho_d$.

What will happen when considering the fact that the probability of
getting $\rho_w$ from $\rho_d$ by NRWMs is not one but $P_w$ given
in Eq.~\eqref{Pw}? Under this situation, the efficiency of the above
entanglement distribution scheme, with NRWMs being performed in
advance on each copy of $\rho_d$, should be
\begin{equation}
 E_f(\rho_d)=P_wE'_f(\rho_w).
\end{equation}
That is, the final efficiency is the product of the efficiencies of
two stages: filtering and distillation protocol. If one does not
recycle the state $\rho_{r_1}$ corresponding to the aforementioned
measurement outcome ``00'', the efficiency
$E_f(\rho_d)=E^1_f(\rho_d)=P_wP_1/2$. It is easy to prove that
$E^1_f(w_1=w_2=0)>E^1_f(w_1\neq 0,w_2\neq 0)$ for arbitrarily given
$d_1$ and $d_2$. This means that the efficiency of the distillation
scheme with NRWM is lower than that of the scheme without NRWM. Such
a conclusion is still tenable in the case of any $m$ rounds of
distillation. As an example, we plot the efficiency $E_f(\rho_d)$
for $m=10$ in Fig.~1. It can be seen from Fig.~1 that $E_f(\rho_d)$
takes the maximum only when $w_1=w_2=0$ (that means no weak
measurement) for any $d_1$ and $d_2$. All the above results imply
that the NRWM does not increase but decreases the efficiency of
bipartite entanglement distribution, i.e., the efficiency of
extracting the Bell state $|\psi\rangle$ from the amplitude-damped
state $\rho_d$. Thus, the NRWM would generate a negative impact on
the bipartite entanglement distribution. The same conclusion will be
obtained for the bisection EDP as shown in the next subsetion.

\begin{figure}[ht]
\centering
\begin{minipage}{8cm}\resizebox{0.9 \textwidth}{!}{%
\includegraphics{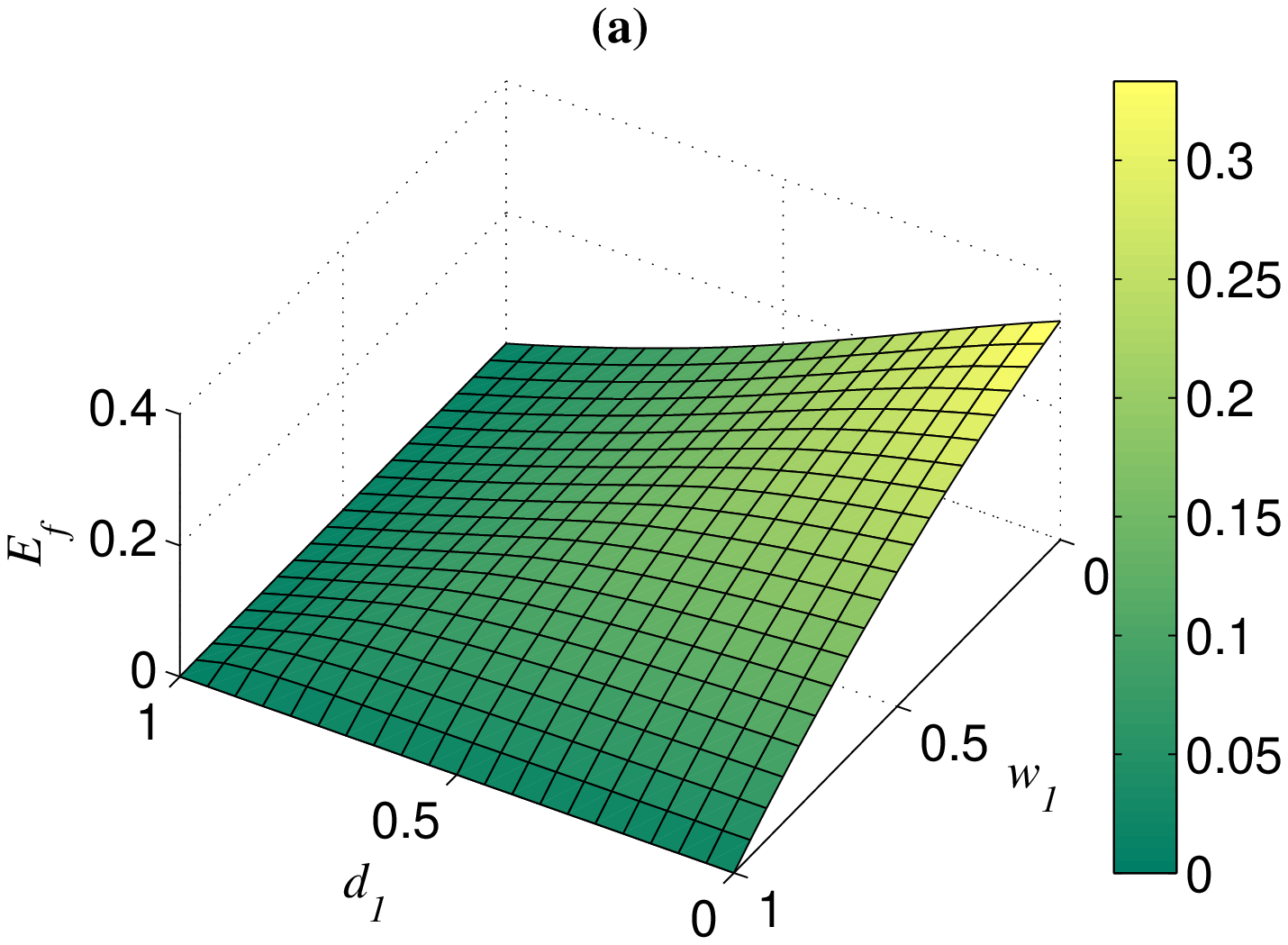}}\end{minipage}
\begin{minipage}{8cm}\resizebox{0.9 \textwidth}{!}{%
\includegraphics{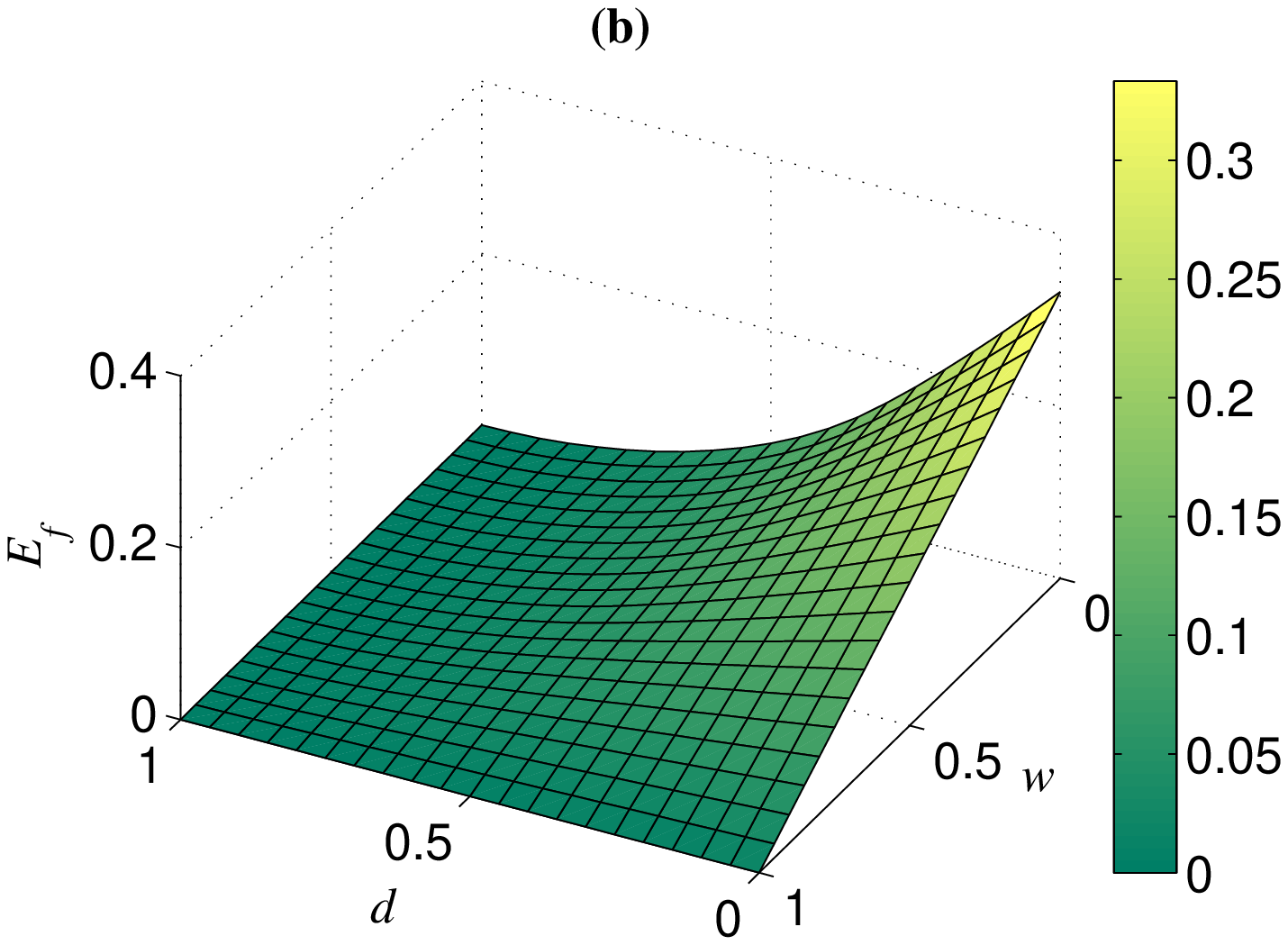}}\end{minipage}
\begin{minipage}{8cm}\resizebox{0.9 \textwidth}{!}{%
\includegraphics{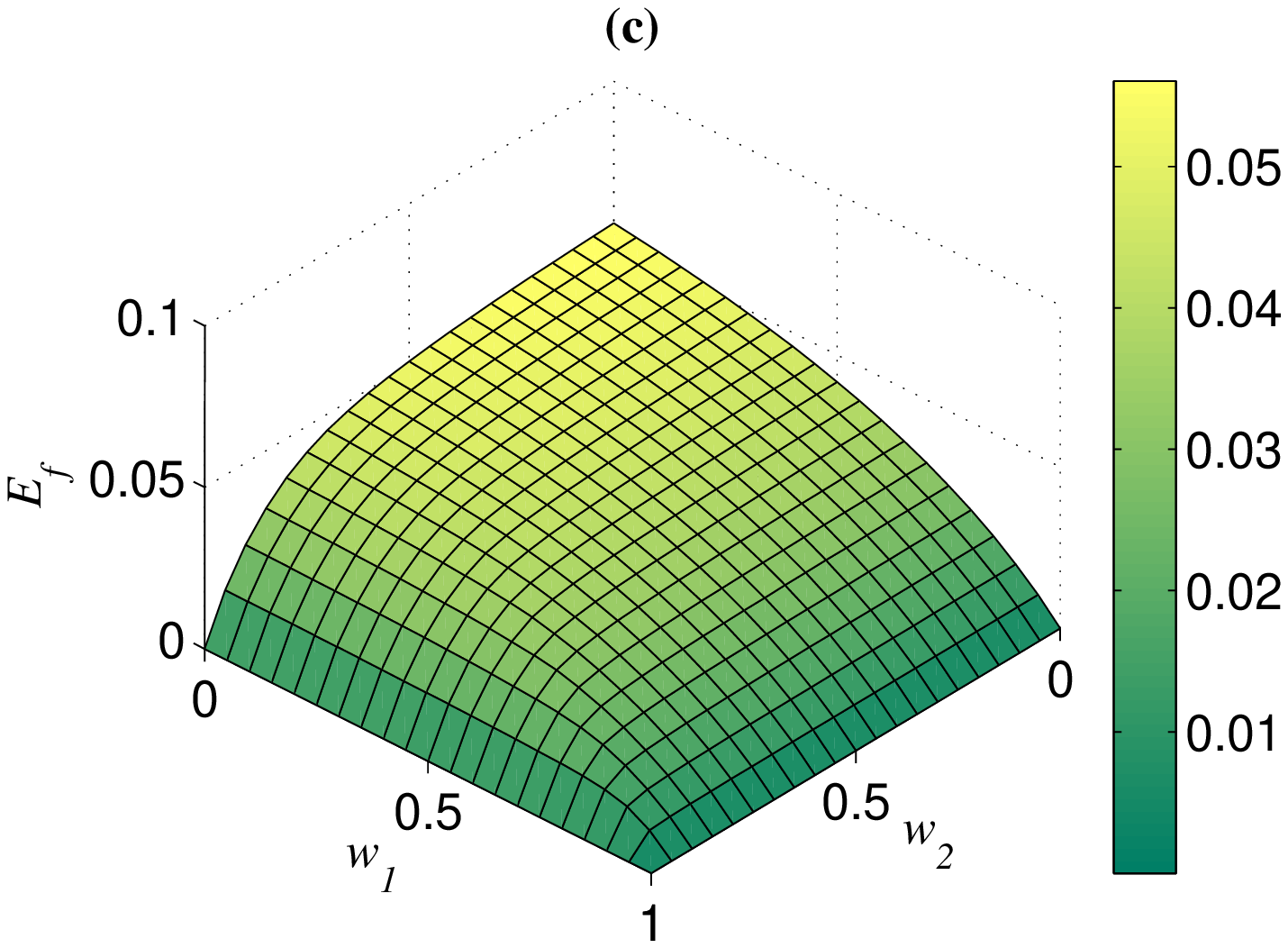}}\end{minipage}
\begin{minipage}{8cm}\resizebox{0.9 \textwidth}{!}{%
\includegraphics{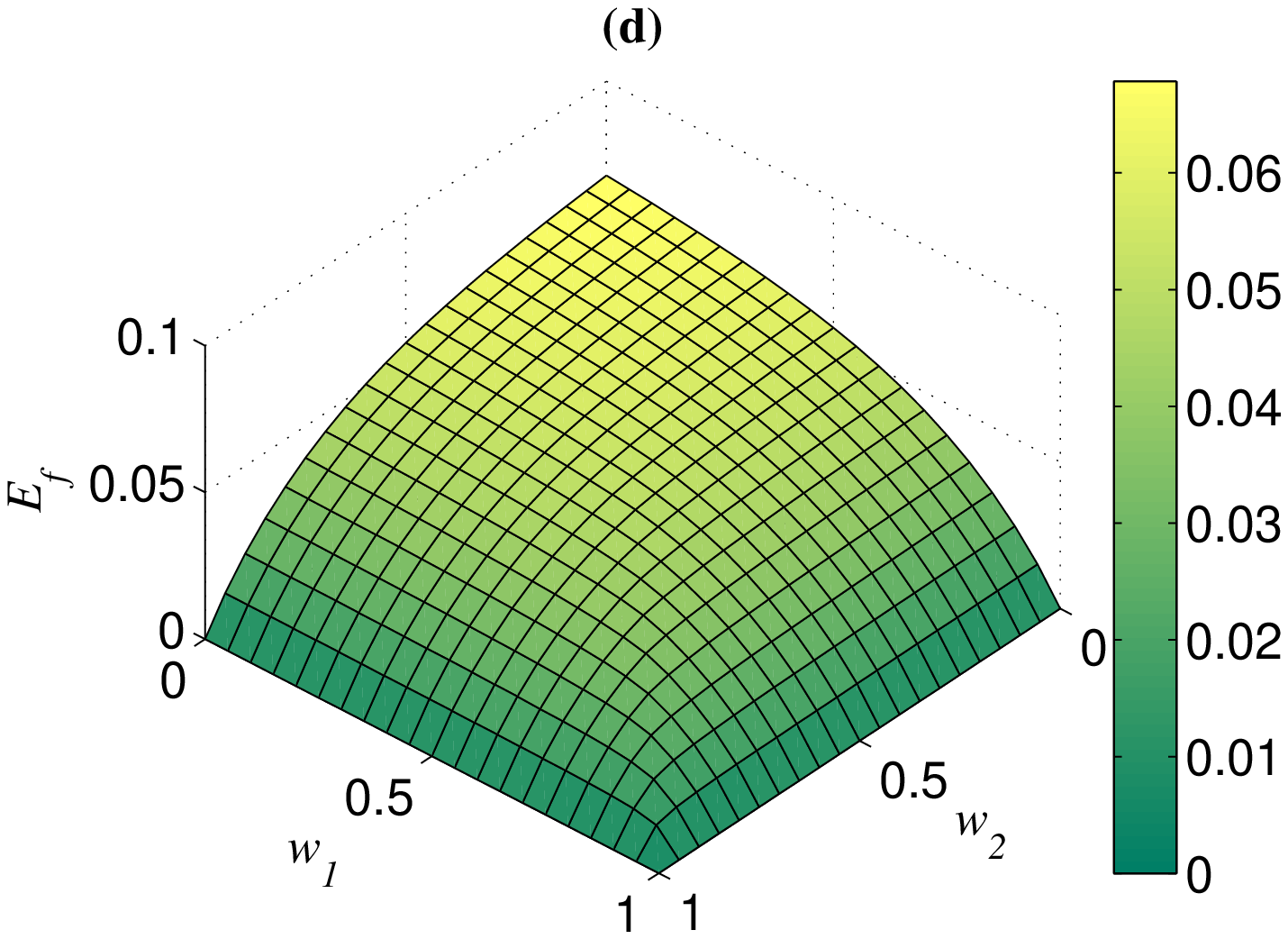}}\end{minipage}
\caption{(Color online) Variation in the bipartite entanglement
distribution efficiency $E_f$ with weak measurement strengths
$(w_1,w_2)$ and channel damping rates $(d_1,d_2)$. Here the number
of rounds is taken $m=10$. \textbf{(a)} $d_2=0=w_2$; \textbf{(b)}
$d_1=d_2=d$ and $w_1=w_2=w$; \textbf{(c)} $d_1=0.3$ and $d_2=0.7$;
\textbf{(d)} $d_1=d_2=0.5$.} \label{fig1}
\end{figure}

We notice that some 2-qubit partially entangled pure states are more
robust than the Bell state $|\psi\rangle$ in terms of the singlet
fractions of the decohered states when just one qubit interacts with
the AD channel \cite{86PRA020304,90PRA052304}. In this case, the
efficiency of establishing maximally entangled states between Alice
and Bob may be slightly improved by substituting the input Bell
state $|\psi\rangle$ for an appropriate 2-qubit partially entangled
pure state. However, it will make no difference to the conclusion
that the NRWM would reduce the efficiency of preparing nonlocal Bell
states. As an example, we replace the initial Bell state
$|\psi\rangle$ by the nonmaximally entangled pure state
$|\psi'\rangle=\frac{1}{\sqrt{2-d}}|01\rangle+\sqrt{\frac{1-d}{2-d}}|10\rangle$
from which the maximum singlet fraction is obtained when only one
qubit suffers from the AD noise \cite{86PRA020304,90PRA052304}. By
the same procedure as before and setting $d_2=0$ (or $d_1=0$), we
obtain the final efficiency of establishing Bell states between
Alice and Bob, as displayed in Fig.~2. Figure 2 indicates that the
no-NRWM-scheme ($w=0$) still outperforms the NRWM-scheme ($w>0$)
even using $|\psi'\rangle$ as the initial state.

\begin{figure}[ht]
\centering
\resizebox{0.5 \textwidth}{!}{%
\includegraphics{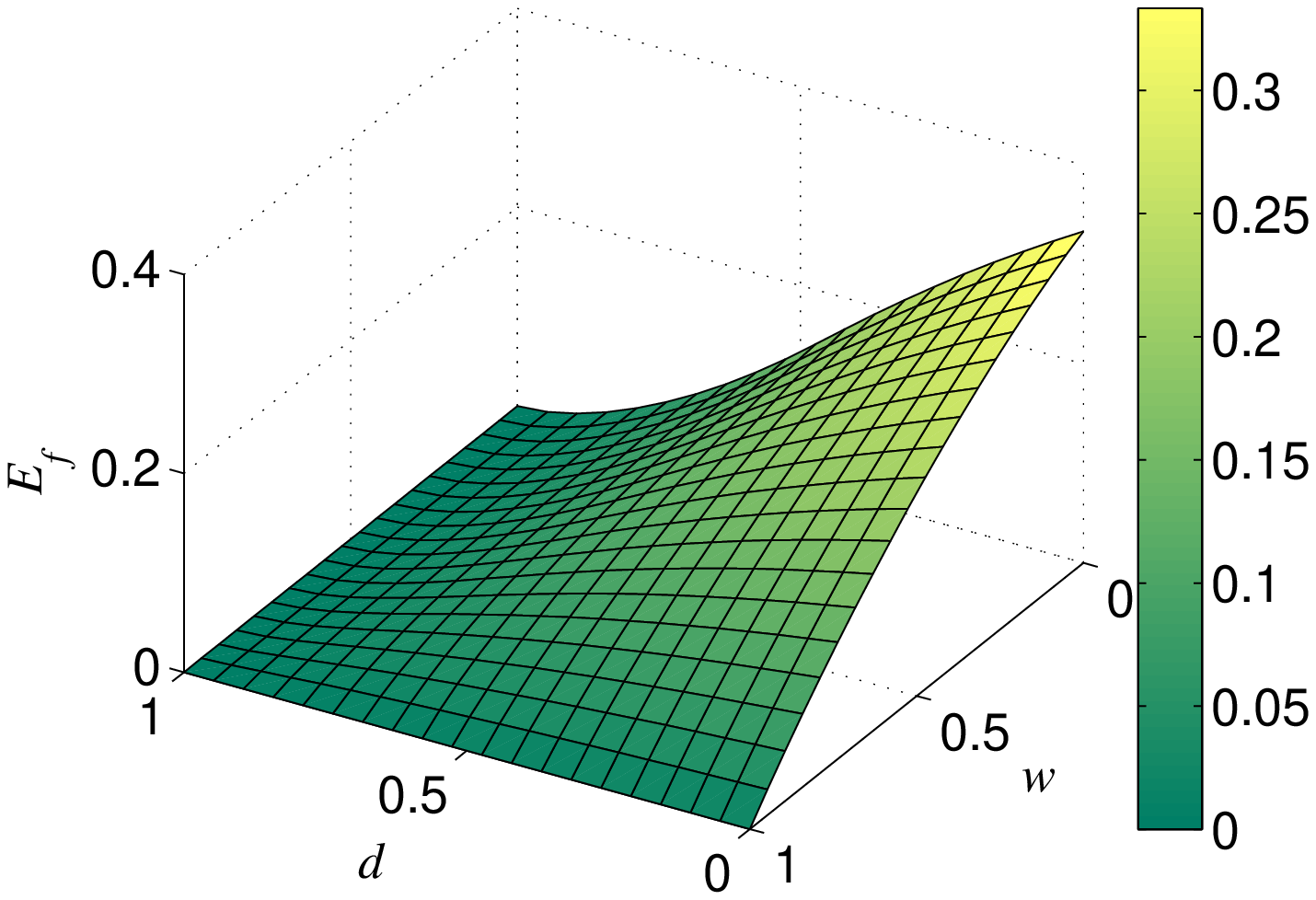}}
\caption{(Color online) Dependence of the bipartite entanglement
distribution efficiency $E_f$ on the weak measurement strength $w$
and channel damping rate $d$ when using
$|\psi'\rangle=\frac{1}{\sqrt{2-d}}|01\rangle+\sqrt{\frac{1-d}{2-d}}|10\rangle$
as the initial state and just one qubit of it suffers from the AD
noise.} \label{fig2}
\end{figure}

\subsection{Bisection EDP}

Let Alice and Bob share $n$ copies of state $\rho_w$, where $n$ is
the power of two. For simplicity, we assume $d_1=d_2=d$ and
$w_1=w_2=w$. Then $\rho^{\otimes n}_w$ can be conveniently written
as
\begin{eqnarray}
 \rho^{\otimes
 n}_w&=&t^n|\psi\rangle\langle\psi|^{\otimes n}
     +t^{(n-1)}\bar{t}\left[|\psi\rangle\langle\psi|^{\otimes
     (n-1)}|00\rangle\langle00|+\cdots\right]+\nonumber\\
     &&t^{(n-2)}\bar{t}^2\left[|\psi\rangle\langle\psi|^{\otimes (n-2)}|00\rangle\langle00|^{\otimes2}+\cdots\right]
     +\cdots \bar{t}^n|00\rangle\langle00|^{\otimes n},
\end{eqnarray}
where
\begin{equation}
\label{t}
  t=\frac{\bar{d}}{d\bar{w}+\bar{d}},
\end{equation}
$\bar{t}=1-t$, and ``$\cdots$'' in each square bracket denotes all
permutations of the first term in the square bracket.

Now both Alice and Bob project her/his part of the state
$\rho^{\otimes n}_w$ on a subspace spanned by vectors with definite
number of ``1''. That is, they perform their particles von Neumann
measurements given by the sets of projectors
\begin{eqnarray}
 &&\left\{M_a^n=\sum\limits_{[x_a]=n,|x_a|=a}|x_a\rangle\langle
   x_a|\right\},\\
&&\left\{M_b^n=\sum\limits_{[y_b]=n,|y_b|=b}|y_b\rangle\langle
   y_b|\right\},
\end{eqnarray}
respectively, where $|x_a|$ ($|y_b|$) denotes the Hamming weight of
the string $x_a$ ($y_b$) of $[x_a]=n$ ($[y_b]=n$) qubits and
$a,b\in\{0,1,\cdots,n\}$. If Alice obtains the measurement outcome
$M_a^n$ and Bob obtains $M_b^n$, the state of the $n$ pairs
collapses into
\begin{eqnarray}
 \rho^{n,n}_{a,b}=\frac{1}{p(n,a,b)}\sum\limits_{[x_a]=[y_a]=n,|x_a|=|y_a|=a}^{[x_b]=[y_b]=n,|x_b|=|y_b|=b}
  |x_a\rangle\langle y_a|\otimes |x_b\rangle\langle y_b|
\end{eqnarray}
with $|x_a\oplus x_b|=|x_{a+b}|$ and $|y_a\oplus y_b|=|y_{a+b}|$,
where ``$\oplus$'' standing for modulo-2 sum of the bitwise of two
strings $x_a$ ($y_a$) and $x_b$ ($y_b$), e.g., $1000+0100=1100$. The
sign ``$\otimes$'' in the above equation denotes the partition
``Alice:Bob'' of $2n$ qubits and $p(n,a,b)$ is given by
\begin{eqnarray}
 p(n,a,b)=\left(
\begin{array}{c}
n\\
a+b
\end{array}
 \right)
\left(
\begin{array}{c}
a+b\\
a
\end{array}
\right).
\end{eqnarray}
The probability of this event is
\begin{equation}
  P(n,a,b)=2^{-a-b}t^{a+b}\bar{t}^{n-a-b}p(n,a,b).
\end{equation}

If $a+b=n$, then Alice and Bob share a maximally entangled pure
state of the rank
\begin{eqnarray}
 r^n_a=\left(
\begin{array}{c}
n\\
a
\end{array}
 \right),
\end{eqnarray}
which is equivalent to $\log_2r^n_a$ maximally entangled pairs of
qubits. If any one of the equalities $\{a=0, a=n, b=0 ,b=n\}$ holds,
Alice and Bob share a separable state. In the remaining cases, the
state $\rho^{n,n}_{a,b}$ is inseparable in terms of the partition
``Alice:Bob'', that is reusable in the second round of distillation.
Using the bisection method (Alice and Bob divide the pairs of qubits
into two blocks of equal length) in the following rounds of
distillation, the total yield of such an EDP starting from the state
$\rho_w$ is given by \cite{80PRA014303}
\begin{eqnarray}
 E'_s(\rho_w)=\sum\limits_{k=1}^{m}t^{2^k}\left[H(2^k)-H(2^{k-1})\right],
\end{eqnarray}
where $m=\log_2n$ and
\begin{eqnarray}
 H(x)=\frac{1}{x2^x}\sum\limits_{l=0}^x\left(
  \begin{array}{c}
   x\\
   l
  \end{array}
 \right)\log_2\left(
  \begin{array}{c}
   x\\
   l
  \end{array}
 \right).
\end{eqnarray}
Considering the fact that the probability of obtaining $\rho_w$ from
the original decohered state $\rho_d$ is $P_w$ as Eq.~\eqref{Pw}
with $d_1=d_2=d$ and $w_1=w_2=w$, the final efficiency of Alice and
Bob sharing maximally entangled pure states should be
\begin{equation}
 E_s(\rho_d)=P_wE'_s(\rho_w).
\end{equation}
The efficiency $E_s(\rho_d)$ as a function of $d$ and $w$ with
$n=32$ is exhibited in Fig.~3. It can be seen that $E_s(\rho_d)$
takes the maximum only when $w=0$ (that means no weak measurement)
for an arbitrarily given $d$, and that the larger $w$, the lower
$E_s(\rho_d)$. This result further justifies the fact that NRWMs
would decrease the efficiency of distributing maximally entangled
pairs to two distant parties. Note that although the total yield of
the bisection EDP could be further improved by combining one-way
hashing method after the first round of distillation
\cite{80PRA014303}, it will not change the conclusion above, due to
the fact that the yields of all rounds except the first round of
distillation procedure are not related with the weak measurement
parameter $w$. In addition, we can see from Fig.~1 and Fig.~3 that
the decrease of $E_s(\rho_d)$ caused by NRWMs in the bisection
protocol is larger than that in the two-copy protocol. It implies
that the more efficient the EDP is, the larger adverse impact the
NRWM will have.

\begin{figure}[ht]
\centering
\resizebox{0.5 \textwidth}{!}{%
\includegraphics{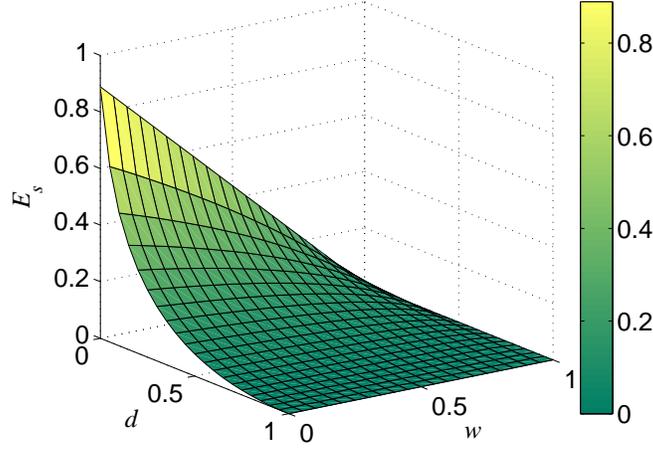}}
\caption{(Color online) Variation in the bipartite entanglement
distribution efficiency $E_s$ with $d$ and $w$. Here we take the
number of source pairs $n=32$.} \label{fig3}
\end{figure}

The negative influence of the NRWM on the above-mentioned bipartite
entanglement distribution could be partly understood from that as
follows. If putting the source states (original noisy states)
through local filters prior to starting distillation procedure, then
the final entanglement distribution efficiency is the product of the
efficiencies of two stages: filtering and distilling. Although the
NRWM could increase the yield of the second stage, it will decrease
the success probability of the first stage (the probability is one
when no weak measurement is performed). The competition of two
opposite effects in two stages leads to the result above.

What is the case for multipartite entanglement distribution? In the
next section, we will elucidate such a problem by discussing the
impact of the NRWM on GHZ-state and W-state distributions,
respectively.

\section{Multipartite entanglement distribution}

\subsection{Distribution of GHZ states}

In this section, we investigate the effect of the NRWM on the
efficiency of GHZ-state distribution in the AD environment based on
multipartite EDPs. The existing GHZ-state distillation protocols
only deal with ``Werner-type'' or GHZ-diagonal states and work in
asymptotic ways \cite{57PRA4075,7QIC689,71PRA012319,74PRA052316}. It
is not clear whether these protocols can be applied to
amplitude-damped GHZ states which are not GHZ-diagonal states. We
here present an efficient GHZ-state distillation protocol which is
suitable for the scenario considered here. This protocol works out
of asymptotic way, and can be regarded as a generalization of the
aforementioned two-copy EDP for two qubits to a multipartite case.
For clarity and simplicity, we here just discuss the case of 3-qubit
GHZ-state distillation and distribution, and the obtained results
can be extended to $N$-qubit GHZ states.

Suppose the initial 3-qubit GHZ state to be distributed to three
distant parties is in the form
\begin{equation}
\label{GHZ}
   |GHZ\rangle=\frac{1}{\sqrt{2}}(|001\rangle+|110\rangle),
\end{equation}
where the three qubits are not all parallel. After each qubit
independently suffering the AD decoherence during the process of
distribution or storage, the state $|GHZ\rangle$ will degrade into
an entangled mixed state denoted by $\rho'_d$. If assume the
decoherence strength of every qubit is the same, the noisy GHZ state
is in the form
\begin{eqnarray}
\label{noisyGHZ}
& \rho'_d=&\frac{1}{2}\left[d(1+d)|000\rangle\langle 000|+\bar{d}|001\rangle\langle 001|+d\bar{d}|010\rangle\langle 010|\right.\nonumber\\
& & \left.+d\bar{d}|100\rangle\langle 100|
+\bar{d}^2|110\rangle\langle 110|
+\sqrt{\bar{d}^3}|001\rangle\langle 110|
+\sqrt{\bar{d}^3}|110\rangle\langle 001|\right].
\end{eqnarray}
We now perform each qubit a NRWM described by the operator $M_w$
given in \eqref{M}. Under the successful event, the noisy GHZ state
becomes $\rho'_w$ which can be obtained by multiplying each
`$|0\rangle$' or `$\langle 0|$' of $\rho'_d$ by the factor
$\sqrt{\bar{w}}$
 (e.g.,$|000\rangle\langle 000|\rightarrow \bar{w}^3|000\rangle\langle 000|$). The success probability is
\begin{equation}
P'_w=\frac{\bar{w}}{2}\left[d(1+d)\bar{w}^2+\bar{d}\bar{w}+2d\bar{d}\bar{w}+\bar{d}^2\right].
\end{equation}
According to the analysis in Ref.~\cite{350AP50}, $\rho'_w$ could be
more entangled than $\rho'_d$ in terms of the measures of negativity
and multipartite concurrence, and thus the fidelity of the former
could also be higher than that of the latter \cite{89PRA062320}.

However, we shall show that the NRWM is not good for distilling pure
GHZ states from noisy GHZ states. The proposed distillation protocol
works as follows: (i) All the three parties take two copies of the
input state $\rho'_w$ (or $\rho'_d$ with $w=0$); (ii) each one
labels the first qubit control and the second target and perform a
CNOT-gate operation on his/her two qubits; (iii) they measure their
target qubits in the basis $\{|0\rangle,|1\rangle\}$; (iv) they keep
the control qubits if they get the outcome ``111'' (this means the
success of extracting the pure GHZ state $|GHZ\rangle$) or ``000''
(in this case, the control copy can be used for the second round of
distillation), and discard it otherwise. Following the same
processing as the bipartite two-copy EDP introduced above, we obtain
the formula of the final distribution efficiency of the GHZ state
$|GHZ\rangle$,
\begin{eqnarray}
 &&E=P'_w\left(Y_1+Y_2+\cdots+Y_m\right),\\
 &&Y_1=\frac{\bar{d}^3\bar{w}^3}{4(P'_w)^2},\nonumber\\
 &&Y_m=\frac{(\bar{d}\bar{w})^{3\cdot 2^{m-2}}Y_{m-1}}{2\left\{\left[d(1+d)\bar{w}^3\right]^{2^{m-1}}+\left(\bar{d}\bar{w}^2\right)^{2^{m-1}}+2\left(d\bar{d}\bar{w}^2\right)^{2^{m-1}}+\left(\bar{d}^2\bar{w}\right)^{2^{m-1}}\right\}},~~(m>1),\nonumber
\end{eqnarray}
where $m$ denotes the number of rounds. The specific dependence  of
the efficiency $E$ on the parameters $d$ and $w$ for $m=10$ is
exhibited in Fig. 4. From figure 4, we can see that $E$ takes the
maximum value only when $w=0$ (that means no weak measurement) for
any $d$. This result means that the NRWM is bad for the distribution
of the 3-qubit GHZ state. We believe the conclusion is also
applicable to $N$-qubit GHZ states. The origin of the negative
influence of the NRWM on the GHZ-state distribution may be the same
as that of Bell-state distribution.

\begin{figure}[ht]
\centering
\resizebox{0.5 \textwidth}{!}{%
\includegraphics{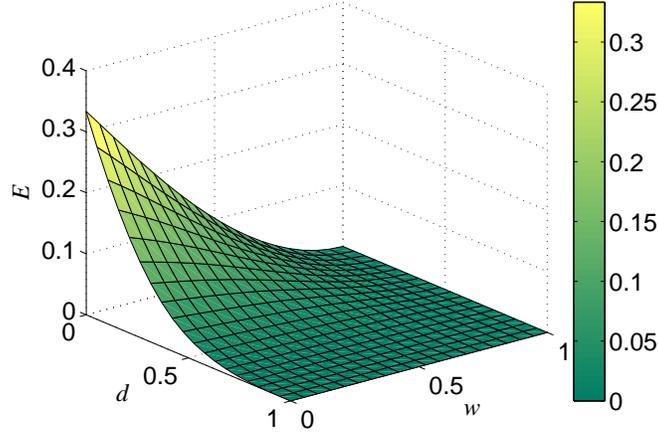}}
\caption{(Color online) Dependence of the 3-qubit GHZ-state
distribution efficiency $E$ on the weak measurement strength $w$ and
channel damping rate $d$.} \label{fig4}
\end{figure}

\subsection{Distribution of W states}

Next, we discuss the role of the NRWM in the distribution of W
states and show different phenomena from that observed before. The W
state is a peculiar type of multipartite entangled state, and has
attracted particular interest on its properties and applications
\cite{62PRA062314,63PRA020303,8QIP319,79PRA062315,283OC4113,74PRA062320,74PRA054303}.
Entanglement distillation of 3-qubit dephased and depolarized W
states was studied in Ref.~\cite{95PRL220501}. A EDP for the 3-qubit
amplitude-damped W state has been proposed in
Ref.~\cite{61JKPS1938}. Here, we show that the EDP in
Ref.~\cite{61JKPS1938} can be generalized to $N$-qubit W states, and
that the yields of W-state distillation schemes could be improved by
the aforementioned NRWM. Moreover, the fidelity thresholds for
distillability of decohered W states could be reduced to near zero.

Suppose the perfect $N$-qubit W state
\begin{equation}
  |W_N\rangle=\frac{1}{\sqrt{N}}(|0\cdots 01\rangle_N+|0\cdots 010\rangle_N+\cdots+|10\cdots0\rangle_N)
\end{equation}
is distributed to $N$ parties (Alice, Bob, Charlie, $\cdots$), but
suffers typical decoherence as described by the local AD channel
with the same damping rate $d$. Then the $N$ parties initially share
a noisy W state given by
\begin{equation}
 \varrho_d=\bar{d}|W_N\rangle\langle W_N|+d|0\cdots0\rangle\langle 0\cdots0|.
\end{equation}
The fidelity of this noisy W state relative to the original pure W
state is $F=\bar{d}$.

We show that the fidelity of $\varrho_d$ can be improved
probabilistically by performing each qubit a NRWM described in
Eq.~\eqref{M}. After each of the $N$ parties performing a NRWM on
the qubit he/she holds with a successful event, the state
$\varrho_d$ becomes
\begin{eqnarray}
  \varrho_w=F_w|W_N\rangle\langle W_N|+\bar{F}_w|0\cdots0\rangle\langle0\cdots0|,
\end{eqnarray}
where the fidelity $F_w=\bar{d}/(d\bar{w}+\bar{d})$ is the same with
$t$ in Eq.~\eqref{t} and $\bar{F}_w=1-F_w$. The success probability
is
\begin{equation}
 p_w=\bar{w}^{N-1}\left(\bar{d}+d\bar{w}\right).
\end{equation}
Obviously, $F_w>F$ as long as the weak measurement strength $w\neq
0$. Thus the fidelity of a single copy of noisy W state $\varrho_d$
can be indeed enhanced by NRWMs by sacrificing a reduction in the
probability. It is easy to see that when $F_w\rightarrow 1$
(corresponding to $w\rightarrow 1$), the success probability
$p_w\rightarrow 0$.

We now demonstrate that the NRWM can improve the efficiency of
distributing the $N$-qubit W state $|W_N\rangle$ in the AD
environment by employing the EDP for amplitude-damped W states.
Suppose there are many groups of $N$-qubit amplitude-damped W states
$\varrho_w$. Each group contains two copies, one as the control copy
and the other as the target copy. $N$ qubits of each copy belong to
$N$ users (Alice, Bob, Charlie, $\cdots$), respectively. The W-state
distillation protocol is as follows: (i) the $N$ users first
perform, respectively, a local CNOT gate between two copies of each
group, the control copy consists of the $N$ control qubits and the
target one the $N$ target qubits; (ii) they then measure locally the
qubits of the target copy in the computational basis
$\{|0\rangle,|1\rangle\}$; (iii) they keep the control copy if they
get the measurement outcome ``$00\cdots 0$'', and discard it
otherwise. Depending on the outcome ``$00\cdots 0$'' known through
classical communication, the $N$ parties share another entangled
mixed state
\begin{eqnarray}
 \varrho_{1}=F_{1}|W_N\rangle\langle W_N|+\bar{F}_1|0\cdots0\rangle\langle0\cdots0|,
\end{eqnarray}
where the fidelity $F_1$ of the noisy W state after the first step
of distillation is given by
\begin{equation}
\label{F1}
 F_{1}=\frac{F^2_w}{F^2_w+N\bar{F}_w^2}.
\end{equation}
The success probability is
\begin{equation}
\label{p1}
 p_{1}=\frac{1}{N}F^2_w+\bar{F}_w^2.
\end{equation}
It is easy to prove that $F_{1}>F_w$ for $F_w>N/(N+1)$. If $w=0$,
meaning that no NRWM has been performed prior to the distillation
operations, only $d<1/(N+1)$ can ensure $F_{1}>F_w(w=0)=F$. It
indicates that the EDP does not work if one directly use the
decohered state $\varrho_d$, instead of $\varrho_w$ ($w>0$), as the
input of it when $d\geqslant 1/(N+1)$. As to the case of $w>0$,
however, the condition of $F_{1}>F_w$ (i.e., $F_w>\frac{N}{N+1}$) is
$d<1/(N\bar{w}+1)$. Evidently, the upper bound of $d$ in this case
could be close to unit by modulating $w$ to be near to one. In other
words, for any damping rate $d$, the NRWM would enable the above EDP
to work, at least in principle, by meeting
\begin{equation}
\label{wvalue}
  w>\frac{(N+1)d-1}{Nd}.
\end{equation}
Note that the degree of weak measurement $w$ can take any value from
0 to 1, and that the inequality \eqref{wvalue} naturally holds for
$d<1/(N+1)$. Moreover, when $d\geqslant 1/(N+1)$, the larger $d$ is,
the larger $w$ is required for satisfying the inequality
\eqref{wvalue}.

So, for the case of $d\geqslant 1/(N+1)$, the NRWM is evidently
beneficial to the distribution of the $N$-qubit W state
$|W_N\rangle$, due to the fact that it can decrease the fidelity
threshold for distillability of the decoehred W state $\varrho_d$
from $N/(N+1)$ to an arbitrarily small number. Whether the NRWM
could still bring benefits in the regime of $d<1/(N+1)$ (keeping it
in mind that the EDP can work with the absence of the NRWM under
this case)? We next focus on discussing such a problem. It will be
shown that the NRWM would contribute to raising the efficiency of
the W-state distillation protocol for most values of $d$ even in the
range $(0,\frac{1}{N+1})$, which indicates that the entanglement
distribution scheme with NRWM could outperform the scheme without
NRWM in most region of $d\in(0,\frac{1}{N+1})$.

Based on the success of the first distillation step, the users can
carry out the second recurrence step by using $\varrho_{1}$ as the
input state. By the same token, they can carry on with the third,
the fourth, and up to the $m$th recurrence step so that obtaining
the nearly perfect W state. In each step, the input states are the
states that are kept in the former step with successful events. The
fidelity and success probability in each step comply with the
recursion formulas \eqref{F1} and \eqref{p1} with $F_w$ being
substituted by the fidelity in the former step. Then after $m$
steps, the fidelity $F_m$ of the obtained state relative to the
initial perfect W state $|W_N\rangle$ and the corresponding
efficiency $E_m^{(N)}$ are given by
\begin{eqnarray}
\label{Fm}
 F_m&=&\frac{1}{1+\lambda_m},\\
 \label{Em}
 E_m^{(N)}&=&p_w\cdot\prod\limits_{i=1}^m\frac{p_i}{2}
    =\bar{w}^{N-1}\bar{d}(1+\lambda_m)\prod\limits_{i=0}^{m-1}\frac{1}{2N(1+\lambda_i)},\\
  p_i&=&\frac{1+\lambda_i}{N(1+\lambda_{i-1})^2}~~(i=1,2,\cdots,m),\nonumber\\
 \lambda_i&=&\frac{1}{N}\left(N\bar{w}\frac{d}{\bar{d}}\right)^{2^i}~~(i=0,1,\cdots,m).\nonumber
\end{eqnarray}
Here $p_i$ denotes the success probability in the $i$th step. If the
fidelity $F_m\geqslant 1-\epsilon$, it means the users obtain a
nearly perfect W state denoted by $|W^{\epsilon}_N\rangle$ and the
W-state distribution succeeds. Following the iteration process as
described above, the distribution of the $N$-qubit W state would be
accomplished in several steps with finite copies of noisy W state
$\varrho_d$.

\begin{figure}[ht]
\centering \label{fig5}
\resizebox{0.5 \textwidth}{!}{%
\includegraphics{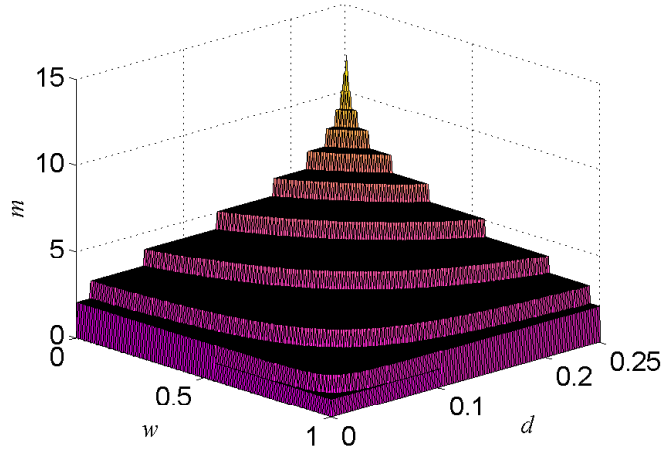}}
\caption{(Color online) The number of distillation steps $m$ needed
for finally getting the aim state $|W_3^{\epsilon_0}\rangle$.}
\end{figure}

\begin{figure}[ht]
 \centering
 \label{fig6}
\resizebox{0.5 \textwidth}{!}{%
\includegraphics{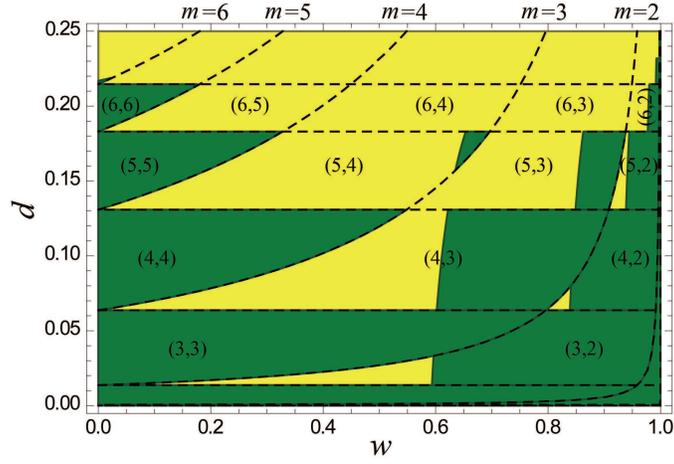}}
\caption{(Color online) The ranges of $d$ and $w$ in which
$R^{(3)}>1$ (yellow) or $\leqslant 1$ (green) under the fidelity
threshold $1-\epsilon_0$. The dashed curve lines and straight lines
correspond, respectively, to $3\bar{w}d/\bar{d}=C^{2^{-m}}$ and
$3d/\bar{d}=C^{2^{-m}}$ ($C=3\epsilon_0/(1-\epsilon_0)$) with
different $m$ (see Appendix A). The pair-wise numbers $(m', m)$
($m'\geqslant m$) denote that the no-NRWM-scheme and the NRWM-scheme
involve, respectively, $m'$ and $m$ steps of distillation so that
the final fidelity of the noisy W state exceeds the threshold
$1-\epsilon_0$ in the encircled regions (see Appendix A). Note that
the no-NRWM-scheme involves only the parameter $d$ in the region of
$(m', m)$. }
\end{figure}

We now take $\epsilon=\epsilon_0=10^{-6}$ as an example for detailed
analysis. For clarity, we first consider $N=3$. The required number
of distillation steps $m$ for getting the aim state
$|W^{\epsilon_0}_3\rangle$ is given in Fig.~5 (see also Appendix A).
From figure 5, we can see that for a given $d$, there always exists
a region of $w>0$ in which the required distillation steps are less
than that for the case with $w=0$. It means that the NRWM can reduce
the number of the distillation steps for obtaining the same expected
state. The step-wise behavior in Fig.~5 implies that to arrive at
the given fidelity threshold, those initial fidelities in a certain
region need the same number of iteration steps. This is due to the
fact that a smaller initial fidelity may lead to a faster increase
in fidelity of the distilled state, which should result from
nonlinearity of the iteration formula of fidelity (given in
Eq.~\eqref{F1}) and the initial fidelity $F_w(d,w)$ with respect to
$d$ and $w$. The advantage  of the NRWM-scheme in distillation steps
can not ensure its efficiency being higher than that of the
no-NRWM-scheme. To judge whether the NRWM-scheme could be superior
to the no-NRWM-scheme, we need to observe the ratio of the
efficiency of the NRWM-scheme $E_m^{(3)}(w\neq 0)$ to that of the
no-NRWM-scheme $E_{m'}^{(3)}(w=0)$, i.e.,
\begin{equation}
R^{(3)}=\frac{E_m^{(3)}(w\neq 0)}{E_{m'}^{(3)}(w=0)},
\end{equation}
Note that $m\leqslant m'$ as shown in Fig.~5. The dependence of
$R^{(3)}$ on $d$ and $w$ is exhibited in Fig.~6, where the region
with denotation $(m', m)$ denotes that the no-NRWM-scheme and the
NRWM-scheme at least involve, respectively, $m'$ and $m$ steps of
distillation so that the final fidelity of the noisy W state exceeds
the threshold $1-\epsilon_0$. It can been seen from Fig.~6 that the
regions of $R^{(3)}>1$ are as follows: (i) $(m'>6,m>2)$; (ii)
$(m'=6,6>m\geqslant 3)$; (iii) most part of $(m'=5, 5>m\geqslant
3)$; (iv) about half of $(4,3)$; (v) part of $(m'\geqslant 3,m=2)$.
It implies that when $m=m'$ with $m'$ being less than a threshold,
$R^{(3)}\leqslant 1$. In other cases, however, the regularity of the
sign of $R^{(3)}-1$ seems to be not clear. It is worth pointing out
that the zig-zag behavior in Fig.~6 is well correspondent with the
step-wise behavior in Fig.~5. Moreover, the non-ordered phenomenon
in Fig.~6 should be related to the fact that $F_m$ is nonlinear with
respect to $d$ and $w$, and that $E_m$ and thus $R^{(3)}$ are
nonmonotonic with respect to $w$. In a word, the ratio $R^{(3)}$
could be larger than one for most values of $d$ in the range
$(0,1/4)$. Thus the NRWM-scheme can indeed outperform the
no-NRWM-scheme in most region of $0<d<1/4$ in terms of the
efficiency. Generally, the larger the degree of decoherence is, the
clearer the superiority of the NRWM-scheme. Moreover, the fact that
the NRWM is helpful to distributing W states does not mean the
larger $w$ the better. The optimal weak measurement strength
$w_o^{(3)}$ that maximizes the efficiency $E_m^{(3)}$ for a given
channel damping rate $d\in(0,1/4)$ is displayed in Fig.~7, where the
inset gives the number of steps $m$ needed for getting the aim state
$|W_3^{\epsilon_0}\rangle$ under the case of $w=w_o^{(3)}$. The jump
phenomenon in Fig.~7 is matching to $R^{(3)}>1$ in the bottom yellow
region of Fig.~6. With the optimal NRWM, we can compute the best
efficiency of 3-qubit W-state distribution $E_o^{(3)}$ (see Fig.~8).

\begin{figure}[ht]
 \label{fig7}
\centering
\resizebox{0.5 \textwidth}{!}{%
\includegraphics{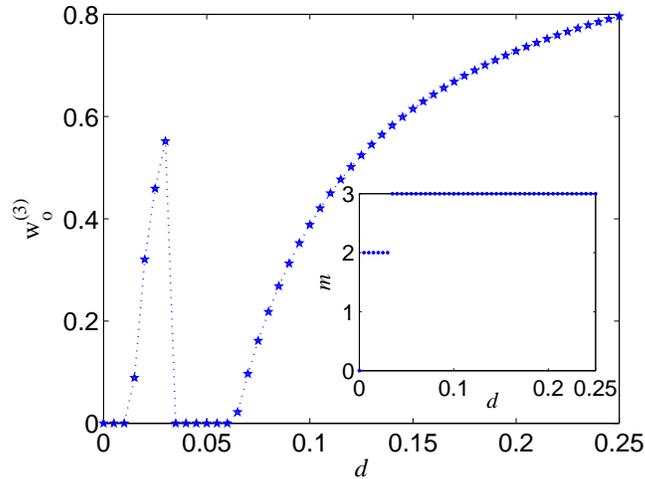}}
\caption{(Color online) The optimal weak measurement strength
$w_o^{(3)}$ that maximizes the efficiency of getting the nearly
perfect W state $|W^{\epsilon_0}_3\rangle$. The inset shows the
required number of distillation steps $m$ for getting
$|W^{\epsilon_0}_3\rangle$ under the optimal degree of weak
measurement $w_o^{(3)}$.}
\end{figure}

\begin{figure}[ht]
 \label{fig8}
\centering
\resizebox{0.5 \textwidth}{!}{%
\includegraphics{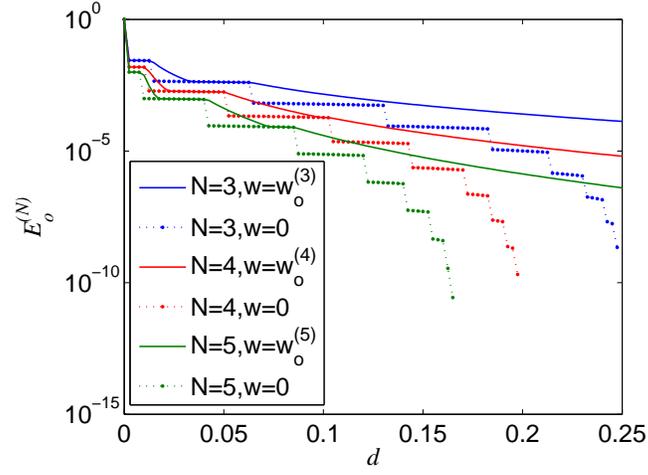}}
\caption{(Color online) The efficiency $E^{(N)}_0$ of distributing
the nearly perfect $N$-qubit W state $|W_N^{\epsilon_0}\rangle$ to
$N$ distant parties in the AD environment. The solid lines denote
the optimal NRWM-scheme and the dotted lines stand for the
no-NRWM-scheme which works only for $d<1/(N+1)$.}
\end{figure}

As for a general $N$, one can still verify that the efficiency of
the EDP with NRWM could be higher than that of the scheme without
NRWM for most values of $d$ in the regime of $d<1/(N+1)$.
Furthermore, we can also find the optimal NRWM strength $w_o^{(N)}$
that maximizes the efficiency of extracting a nearly perfect
$N$-qubit W state $|W_N^{\epsilon_0}\rangle$ from the decohered
state $\varrho_d$ for a given channel damping rate $d$, and then
calculate the corresponding highest efficiency $E_o^{(N)}$. Note
that $w_o^{(N)}$ may be dependent on $N$. As examples, we show the
optimal efficiencies $E_o^{(N)}$ for $N=4,5$ in Fig.~8. It can be
seen that in the regime of $d<1/(N+1)$, the efficiency of the scheme
with NRWM is indeed higher than that of the scheme without NRWM for
most values of $d$.

\begin{figure}[ht]
 \label{fig9}
\centering
\resizebox{0.5 \textwidth}{!}{%
\includegraphics{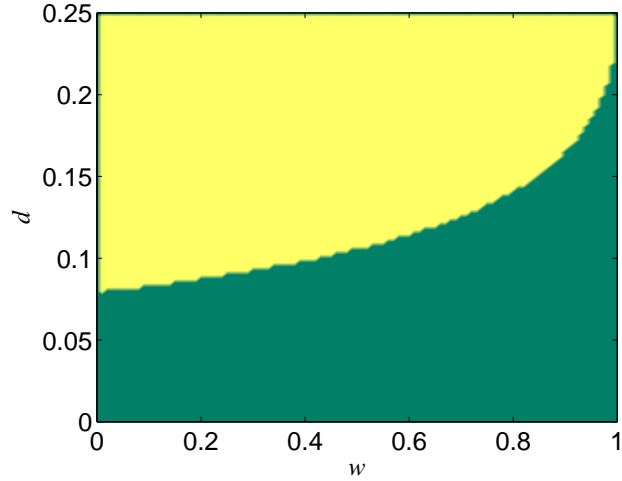}}
\caption{(Color online) The ranges of $d$ and $w$ in which
$\lim\limits_{\epsilon\rightarrow 0}R^{(3)}>1$ (yellow region) or
$\leqslant 1$ (green region).}
\end{figure}

Finally, we give a brief discussion on the case of
$\epsilon\rightarrow 0$. Obviously, $\epsilon\rightarrow 0$ would
lead to the fact that the entailed number of distillation steps
tends to infinity. Then we obtain (see Appendix B)
\begin{eqnarray}
\label{epsilonR}
 \lim\limits_{\epsilon\rightarrow 0}R^{(N)}&=&\lim\limits_{m,m'\rightarrow
\infty}\frac{E_m^{(N)}(w\neq 0)}{E_{m'}^{(N)}(w=0)}\nonumber\\
& \sim
&\bar{w}^{N-1}\exp\left[\frac{1}{\ln2}\int\limits_{N\lambda'_0}^{N\lambda_0}\frac{\ln(2N+2u)}{u\ln
 u}\mathrm{d}u\right],
\end{eqnarray}
where $\lambda'_0=d/\bar{d}$. As an example, we give the ranges of
$d$ and $w$ in which $\lim\limits_{\epsilon\rightarrow 0}R^{(3)}$ is
larger or less than one in Fig.~9. Figure 9 indicates that the ratio
$R^{(3)}$ of the efficiency of the NRWM-scheme to that of the
no-NRWM-scheme could also be larger than one under the case of
$\epsilon\rightarrow 0$ as long as the channel damping rate $d$ is
not too small. In addition, the larger $d$ is, the clearer the
advantage of the NRWM-scheme is. The same results could be obtained
for $N>3$.

The positive impact of the NRWM on the W-state distribution could be
partly explained by the fact that its positive effect in the
distillation phase can surpass its negative effect in the filtering
phase when some conditions are satisfied.

\section{Summary}

Entanglement distillation is a good tool to prepare entangled pure
states among distant parties in noisy environments by concentrating
the entanglement of a large number of decohered states into a small
number of entangled states. Local filtering may be another possible
solution to overcoming decoherence of quantum systems. As claimed, a
particular filter could be utilized to increase the amount of
entanglement of a single-copy noisy entangled state with a ceratin
probability. The filtering method, however, cannot be applied for
direct production of an entangled pure state. The effect of
filtering operations on protecting entanglement from decoherence
would be far more exciting if they can be combined with EDPs to
improve the efficiency of distributing entangled pure states to
distant users who plan to implement remotely faithful quantum
information tasks.

In this paper, we have investigated the possibility of improving the
efficiency of distilling maximally entangled pure states from
entangled mixed states in the AD environment by using the NRWM (a
local filtering operation) which has recently been shown to be an
effective method for enhancing probabilistically the entanglement of
a single-copy amplitude-damped entangled state. We have shown that
NRWMs would lead to the decrease of the distillation efficiencies of
bipartite maximally entangled states and multipartite GHZ states.
However, we found that the NRWM is beneficial to remote
distributions of multipartite W states. We demonstrated that the
NRWM can not only reduce the fidelity thresholds for distillability
of decohered W states, but also raise the distillation efficiencies
of W states. The different effects of the NRWM on the distillation
efficiencies of W and GHZ states (or bipartite maximally entangled
states)  may be related to the fact that the former works in an
asymptotic way but the latter does not.

Our results indicate that the NRWM is not necessarily helpful to
practical entanglement distributions which aim at establishing
maximally entangled pure (or nearly pure) states among distant
parties, although it can increase to some extent the amount of
entanglement of a single-copy entangled mixed state with a certain
probability. This leads to a new criterion for measuring the
usefulness of a local filter in protecting entanglement from
decoherence. These findings are expected to inspire widespread
interest on investigating the possibility of improving efficiencies
of distributing entangled states in noisy environments by local
filtering operations.

\begin{acknowledgements}
This work was supported by the Hunan Provincial Natural Science
Foundation of China (Grant No.~2015JJ3029), the Scientific Research
Fund of Hunan Provincial Education Department of China (Grant
No.~15A028), and also funded by the Singapore Ministry of Education
(partly through the Academic Research Fund Tier 3 Grant
No.~MOE2012-T3-1-009) and the National Research Foundation,
Singapore (Grant No.~WBS: R-710-000-008-271). X. W. Wang was also
supported in part by the Government of China through CSC.
\end{acknowledgements}

\appendix

\section{}

The purpose of the distillation is to make the final fidelity of the
mixed W state $F_m$ reach to the threshold $1-\epsilon$ via the
minimum number of distillation steps $m$. Thus $m$ should satisfies
$F_m\geqslant 1-\epsilon>F_{m-1}~(m\geqslant 1)$. Using
Eq.~\eqref{Fm}, one can readily obtain
\begin{eqnarray}
\label{m1}
 m=\left\lceil\log_2\frac{\ln\frac{N\epsilon}{1-\epsilon}}{\ln(N\lambda_0)}\right\rceil.
 \end{eqnarray}
Next, we take $\epsilon=\epsilon_0$ and $N=3$ for explaining Fig.~6.
With Eq.~\eqref{m1}, we can obtain the equations of dashed curve
lines in Fig. 6,
\begin{eqnarray}
\label{curveline}
  3\bar{w}\frac{d}{\bar{d}}=C^{2^{-m}},~~ C=\frac{3\epsilon_0}{1-\epsilon_0}.
\end{eqnarray}
In the region between neighbored two dashed curve lines with $m$ and
$m-1$, the required number of distillation steps is $m$ for the
NRWM-scheme. The dashed straight lines parallel to $w$ axis can be
directly obtained by setting $w=0$ in Eq.~\eqref{curveline}. Note
that $w=0$ corresponds to the no-NRWM-scheme. So, if
$C^{2^{-(m'-1)}}<3d/\bar{d}\leqslant ^{2^{-m'}}~(m'\geqslant
1)$, the required number of distillation steps is $m'$ for the
no-NRWM-scheme. Then, the region surrounded by neighbored two
straight lines and two curve lines satisfies
$C^{2^{-(m-1)}}<3\bar{w}d/\bar{d}\leqslant C^{2^{-m}}$ and
$C^{2^{-(m'-1)}}<3d/\bar{d}\leqslant C^{2^{-m'}}$, which is denoted
by the pair-wise numbers $(m', m)$ for short. By the way, the
intersection points of curve and straight lines satisfy the equation
$\bar{w}=3d/\bar{d}$. In the region with denotation  $(m', m)$
($m<m'$, as shown in Fig.~5), the entailed numbers of distillation
steps are $m'$ and $m$ for the no-NRWM-scheme and the NRWM-scheme,
respectively. It should be pointed out that $m'=0$ means no
purification operation is needed, and that $m=0$ means purification
task can be accomplished by only weak measurements. In the regions
on and under the curve $3\bar{w}d/\bar{d}=C$, $m=0$. If $d\leqslant
C/(3+C)$, $m'$ is equal to zero and thus $m$ is also equal to
zero. For any given $m'$ and $m$, the boundary of $R^{(3)}\leqslant
1$ can be obtained by solving, at least in principle, the inequality
$E_m\leqslant E_{m'}$, i.e.,
\begin{eqnarray}
  2^{m'-m}x^2\left[3+(xy)^{2^m}\right]\prod\limits_{i=0}^{m'-1}\left(3+y^{2^i}\right)
  \leqslant \left(3+y^{2^{m'}}\right)\prod\limits_{i=0}^{m-1}\left[3+(xy)^{2^i}\right],
\end{eqnarray}
where $x=\bar{w}$ and $y=3d/\bar{d}$. If the inequality has no
solution, it means $R^{(3)}>1$ within the total region $(m',m)$.

\section{}
The derivation of Eq.~\eqref{epsilonR} is given below. When
$m\rightarrow\infty$, we have
\begin{eqnarray}
  &&\prod\limits_{i=0}^{m-1}2N(1+\lambda_i)\rightarrow
  \exp\left[\int\limits_0^{\infty}\ln(2N+2N\lambda_x)\mathrm{d}x\right],\\
  \label{lambdato0}
  && \lambda_m\rightarrow 0,
\end{eqnarray}
where the inequality $N\lambda_0<1$ (because $d<\frac{1}{N+1}$) has
been utilized. Making two times of variable substitutions $y=2^x$
and $u=(N\lambda_0)^y$, one will get
\begin{eqnarray}
\label{int}
  \int\limits_0^{\infty}\ln(2N+2N\lambda_x)\mathrm{d}x=\frac{1}{\ln2}
  \int\limits^{\epsilon}_{N\lambda_0}\frac{\ln(2N+2u)}{u\ln u}\mathrm{d}u,
\end{eqnarray}
where $\epsilon\rightarrow 0$. By substituting
Eqs.~\eqref{lambdato0} and \eqref{int} into Eq.~\eqref{Em}, we
obtain
\begin{eqnarray}
\label{Eminfinity}
 \lim\limits_{m\rightarrow\infty}E_m^{(N)}(w\neq
 0)\approx\bar{w}^{N-1}\bar{d}\exp\left[\frac{1}{\ln2}\int\limits_{\epsilon}^{N\lambda_0}\frac{\ln(2N+2u)}{u\ln u}\mathrm{d}u\right].
\end{eqnarray}
 Similarly, we can obtain
\begin{eqnarray}
\label{Em'infinity}
 \lim\limits_{m'\rightarrow\infty}E_{m'}^{(N)}(w=0)\approx\bar{d}\exp\left[\frac{1}{\ln2}\int\limits_{\epsilon}^{N\lambda'_0}\frac{\ln(2N+2u)}{u\ln
 u}\mathrm{d}u\right],
\end{eqnarray}
where $\lambda'_0=d/\bar{d}$. Eq.~\eqref{epsilonR} can be
straightforwardly derived from Eqs.~\eqref{Eminfinity} and
\eqref{Em'infinity}.

\end{document}